\documentclass [pra,twocolumn,showpacs,dvips]{revtex4}
\usepackage{dcolumn,bm,amsmath}
\usepackage{graphicx}%

\newcommand{\eref}[1]{(\ref{#1})}
\newcommand{\Eref}[1]{Eq.~(\ref{#1})}
\newcommand{\tref}[1]{Table~\ref{#1}}
\newcommand{\rtw}{\rightarrow}
\newcommand{\cm}{cm$^{-1}$}

\begin{document}
\title{The $\bm{\alpha}$-dependence of transition frequencies for ions
Si~II, Cr~II, Fe~II, Ni~II, and Zn~II}
\author{V. A. Dzuba$^1$}
\author{V. V. Flambaum$^1$}
\author{M. G. Kozlov$^2$}
\email{mgk@MF1309.spb.edu}
\author{M. Marchenko$^1$}
\affiliation{$^1$~University of the New South Wales, Sydney, Australia}
\affiliation{$^2$~Petersburg Nuclear Physics Institute, Gatchina, 188300, Russia}
\date{\today}

\begin{abstract}
We performed accurate calculation of $\alpha$-dependence ($\alpha=e^2/hc$)
of the transition frequencies for ions, which are used in a search for the
variation of the fine structure constant $\alpha$ in space-time. We use
Dirac-Hartree-Fock method as a zero approximation and then the many-body
perturbation theory and configuration interaction methods to improve the
results. An important problem of level pseudocrossing (as  functions of
$\alpha$) is considered. Near the crossing point the derivative of
frequencies over $\alpha$ varies strongly (including change of the sign).
This makes it very sensitive to the position of the crossing point.
We proposed a semiempirical solution of the problem which allows
to obtain accurate results.
\end{abstract}


\maketitle


\section{Introduction}

Recently there was an intensive discussion of the possible space-time
variation of the fine structure constant $\alpha=e^2/hc$ at the
cosmological scale. The first evidence for such variation has been reported
in \cite{WFC99,WMF01,MWF01a,MWF01,MWF01c,MWF01d} from the analysis of the
astrophysical data. These results are to be compared with the number of
experimental upper bounds on this variation obtained from other
astrophysical observations (see, e.g. \cite{CW99,IPV99,VIP99}) and from the
precision laboratory measurements \cite{PTM99,SBN99,SBN00}. Recently a
number of new laboratory tests have been proposed (see, e.g. \cite{BPM01}).
The analysis of the microwave background radiation can also give some
restrictions on time variation of $\alpha$ as suggested in
\cite{KST99,Han99,KS00}. Implementations of the space-time variation of the
fine structure constant to the theory of the fundamental interactions are
discussed e.g. in Refs.~\cite{AM97,BIR99,Kuh99,Fuj00,CLV01,KR01,SBM01} (see
also disscussion and references in \cite{MWF01a}).

The most straitforward way to look for the variation of $\alpha$ is to
measure the ratio of some fine structure interval to an optical
transition frequency, such as $\omega( np_{1/2} \rtw np_{3/2})$
and $\omega( n's_{1/2} \rtw
np_{3/2})$~\endnote{In fact, the frequency $\omega( np_{1/2} \rtw
np_{3/2})$ is not measured directly, but is found as a difference:
$\omega(n's_{1/2}\rtw np_{3/2})-\omega( n's_{1/2} \rtw np_{1/2})$.}. This
ratio can be roughly estimated as $0.2\,\alpha^2 Z^2$, where $Z$ is the
nuclear charge \cite{Sob79}. Therefore, any difference in this ratio for a
laboratory experiment and a measurement for some distant astrophysical
object can be easily converted into the space-time variation of $\alpha$.
However, as it was pointed out in \cite{DFW99b}, one can gain about an
order of magnitude in the sensitivity to the $\alpha$-variation by
comparing optical transitions for different atoms. In this case the
frequency of each transition can be expanded in a series in $\alpha^2$:
\begin{subequations}
\label{i1}
\begin{eqnarray}
\label{i1a}
  \omega_i &=& \omega_i^{(0)} + \omega_i^{(2)} \alpha^2 + \dots
\\
\label{i1b}
  &=&
  \omega_{i,\textrm{lab}} + q_i x + \dots,
  \quad x \equiv \left(\alpha/\alpha_0\right)^2-1,
\end{eqnarray}
\end{subequations}
where $\alpha_0$ stands for the laboratory value of the fine structure
constant. Note, that \Eref{i1a} corresponds to the expansion at $\alpha=0$,
while \Eref{i1b} ---~to the expansion at $\alpha=\alpha_0$. In both cases
parameters $\omega^{(2)}_i$ and $q_i$ appear due to relativistic
corrections.

For a fine structure transition the first coefficient on the right hand
side of \eref{i1a} turns to zero, while for the optical transitions it does
not. Thus, for the case of a fine structure and an optical transition one
can write:
\begin{eqnarray}
\label{i2}
  \frac{\omega_{\textrm{fs}}}{\omega_{\textrm{op}}}
  = \frac{\omega_{\textrm{fs}}^{(2)}}{\omega_{\textrm{op}}^{(0)}}
  \, \alpha^2 + O(\alpha^4),
\end{eqnarray}
while for two optical transitions $i$ and $k$ the ratio is:
\begin{eqnarray}
\label{i3}
  \frac{\omega_i}{\omega_k}
  = \frac{\omega_i^{(0)}}{\omega_k^{(0)}}
   +\left(\frac{\omega_i^{(2)}-\omega_k^{(2)}}{\omega_k^{(0)}}
   \right) \, \alpha^2 + O(\alpha^4).
\end{eqnarray}
Quite often the coefficients $\omega_i^{(2)}$ for optical transitions are
about an order of magnitude larger than corresponding coefficients for the
fine structure transitions $\omega_{\textrm{fs}}^{(2)}$ (this is because
the relativistic correction to a ground state electron energy is
substantially larger than the spin-orbit splitting in an excited state
\cite{DFW99a,DFW99b}). Therefore, the ratio \eref{i3} is, in general, more
sensitive to the variation of $\alpha$ than the ratio
\eref{i2}. It is also important that the signs of coefficients
$\omega_i^{(2)}$ in \eref{i3} can vary. For example, for $s$-$p$
transitions the relativistic corrections are positive while for $d$-$p$
transitions they are negative. This allows to suppress possible systematic
errors which ``do not know'' about the signs and magnitude of the
relativistic corrections \cite{DFW99b}. On the other hand, for many cases
of interest, the underlying atomic theory is much more complicated for
\Eref{i3}. In particular, the most difficult case corresponds to
transitions to highly excited states of a multi-electron atom, where the
spectrum is very dense. And this happens to be a typical situation for
astrophysical spectra, in particular, for large cosmological red shifts.
Corresponding atomic calculations have to account very accurately for the
electronic correlations, which may affect such spectra quite dramatically.

The first calculations of the coefficients $q$ from \Eref{i1} for the
transitions suitable for astronomical and laboratory measurements were done
in Refs.~\cite{DFW99a,DFW99b,DF00a,DFM01}. Here we present a new and more
accurate calculations of the coefficients $q$ for the transitions, which
are currently used in the analysis of the astrophysical data. A full list
of these transitions was given in \cite{MWF01a}. We have not recalculated
here the lightest and the most simple atoms Mg and Al, for which the
previous calculation \cite{DFW99b} should be sufficiently accurate and
focused on more complicated ions Si~II, Cr~II, Fe~II, Ni~II, and Zn~II. Our
final results for them are given in \tref{tab_fin}. Note, that here we use
the single parameter $q$ instead of two parameters $q_1$ and $q_2$ used in
the earlier works and $q \equiv \partial
\omega/\partial x|_{x=0} = q_1+2q_2$. Details of the calculations and
discussion of the accuracy will be given in Sec.~\ref{details}.
Before that we briefly address few theoretical points
in Sec.~\ref{theory}.

\begin{table}[tb]
\caption{Final results for parameters $q$ from \Eref{i1} for
Si~II, Cr~II, Fe~II, Ni~II, and Zn~II. Estimated errors are in brackets.}

\label{tab_fin}

\begin{tabular}{llcldrl}
\hline
\hline
 \multicolumn{1}{c}{Ion}
&\multicolumn{3}{c}{Transition}
&\multicolumn{1}{c}{$\omega_0$ (\cm)}
&\multicolumn{2}{c}{$q$ (\cm)}\\
\hline
Si II &$^2\!P^o_{1/2}$&$\rtw$&$^2\!D_{3/2}  $& 55309.3365 &$  520 $& (30)\\
      &               &$\rtw$&$^2\!S_{1/2}  $& 65500.4492 &$   50 $& (30)\\
\hline
Cr II &$^6\!S_{5/2}$&  $\rtw$&$^6\!P^o_{3/2}$& 48398.868  &$-1360 $& (150)\\
      &             &  $\rtw$&$^6\!P^o_{5/2}$& 48491.053  &$-1280 $& (150)\\
      &             &  $\rtw$&$^6\!P^o_{7/2}$& 48632.055  &$-1110 $& (150)\\
\hline
Fe II &$^6\!D_{9/2}$&  $\rtw$&$^6\!D^o_{9/2}$& 38458.9871 &$ 1330 $& (150)\\
      &             &  $\rtw$&$^6\!D^o_{7/2}$& 38660.0494 &$ 1490 $& (150)\\
      &             &  $\rtw$&$^6\!F^o_{11/2}$&41968.0642 &$ 1460 $& (150)\\
      &             &  $\rtw$&$^6\!F^o_{9/2}$& 42114.8329 &$ 1590 $& (150)\\
      &             &  $\rtw$&$^6\!P^o_{7/2}$& 42658.2404 &$ 1210 $& (150)\\
      &             &  $\rtw$&$^4\!F^o_{7/2}$& 62065.528  &$ 1100 $& (300)\\
      &             &  $\rtw$&$^6\!P^o_{7/2}$& 62171.625  &$-1300 $& (300)\\
\hline
Ni II & $^2\!D_{5/2}$& $\rtw$&$^2\!F^o_{7/2}$& 57080.373  &$ -700 $& (250) \\
      &              & $\rtw$&$^2\!D^o_{5/2}$& 57420.013  &$-1400 $& (250) \\
      &              & $\rtw$&$^2\!F^o_{5/2}$& 58493.071  &$  -20 $& (250) \\
\hline
Zn II & $^2\!S_{1/2}$& $\rtw$&$^2\!P^o_{1/2}$& 48481.077  &$ 1584 $& (25) \\
      &              & $\rtw$&$^2\!P^o_{3/2}$& 49355.002  &$ 2490 $& (25) \\
\hline
\hline
\end{tabular}
\end{table}

\section{Theory}
\label{theory}

In order to find parameters $q =\partial \omega/\partial x|_{x=0}$ in
\Eref{i1} we perform atomic calculations for three values of $x$:
$x_-=-1/8$, $x_0=0$, and $x_+=1/8$. That allows us to determine $q$:
$q=4\left(\omega(x_+)-\omega(x_-)\right)$ and also estimate the second
derivative $\partial^2 \omega/\partial x^2|_{x=0}$. The large value of the
latter signals that interaction between levels is strong
(level pseudocrossing), and there is a
risk of large errors. For these cases further analysis was done as
described below.

\paragraph{Relativistic calculations of multi-electron ions.}

In order to accurately account for the dominant relativistic effects we use
the Dirac-Hartree-Fock approximation as a starting point for all
calculations of atomic spectra. Though most of the calculations were done
for the Coulomb potential, we have also estimated Breit corrections by
including the magnetic part of the Breit interaction in the self-consistent
field~\cite{LMY89}.

The ions we are dealing with in this paper have from one to nine electrons
in the open shells. For  one valence electron in Zn II
 the Dirac-Fock $V^{N-1}$
approximation already gives rather good results. On the next step the
core-valence correlations can be accounted for by means of the many-body
perturbation theory (MBPT). Already the second order MBPT correction allows
to reproduce the spectrum with the accuracy, better than 1\%, which is more
than sufficient for our current purposes.

Other ions of interest to us have at least three valence electrons. Here
the dominant correlation correction to transition frequencies corresponds
to the valence-valence correlations. This type of correlations can be
accounted for with configuration interaction (CI) method. If necessary, the
core-valence correlations can be included within combined CI+MBPT technique
\cite{DFK96b}. The latter usually provides an accuracy of the order of 1\%
or better for the lower part of the spectra of atoms and ions with two or
three valence electrons \cite{DFK96b,DJ98,KP99tr}. However, the accuracy of
\textit{ab initio} methods decreases with the number of valence electrons
and with excitation energy. Indeed, for a large number of valence electrons
and/or sufficiently high excitation energy the spectrum becomes dense and
the levels with the same exact quantum numbers strongly interact with each
other. The part of the spectrum of Fe II above 55000 ~\cm\ and, to a
somewhat lesser extent, the spectrum of Ni II represent this situation.
Therefore, for these ions we developed a semiempirical fitting procedure,
which is described below.

In order to have additional control of the accuracy of our CI we performed
calculations for most of the ions with two different computer packages. One
package was used earlier in Refs.~\cite{DFK96b,KP99tr,PKRD01} and another
one was used in Refs.~\cite{DJ98,DFW99a,DFW99b,DF00a,DFM01,MWF01a}. The
former package allows to construct flexible basis sets and optimize
configuration space, while the latter allows for a larger CI space as it
works with the block of the Hamiltonian matrix, which corresponds to a
particular total angular momentum of atom $J$. When there were no
significant difference between two calculations, we only give results
obtained with the first package. Nevertheless, our final results presented
in \tref{tab_fin} are based on both calculations.

\paragraph{Semiempirical treatment of the strong interaction of levels:
pseudo-crossing.} In the nonrelativistic limit $\alpha \rtw 0$, all
multi-electron states are accurately described by the $LS$-coupling scheme:
$E_{\alpha \rtw 0} = E_{p,n,L,S,J}$, where $p=\pm 1$ is the parity and $n$
numerates levels with the same $p$, $L$,$S$, and $J$. For sufficiently
small values of $\alpha$ the $LS$-coupling holds and the energy has the
form:
\begin{eqnarray}
\label{t1}
  &&E_{p,n,L,S,J} = E_{p,n,L,S}^{(0)}
  + \left(\frac{\alpha}{\alpha_0}\right)^2 \biggl(C_{p,n,L,S}
\\ &&
  +  \frac{1}{2}A_{p,n,L,S}\;
  [J(J+1)-L(L+1)-S(S+1)]\biggr),
\nonumber
\end{eqnarray}
where the first term in the parentheses gives the slope for the centre of
the multiplet and the second term gives the fine structure. With growing
$\alpha$ the multiplets start to overlap and when the levels with the same $p$
and $J$ come close, the pseudo-crossing takes place.

Near the pseudo-crossing the slope of the energy curves changes
dramatically. If such crossing takes place at $x \approx 0$, where $x$ is
defined by \Eref{i1}, i.e. near the physical value of $\alpha$, it can
cause significant uncertainty in the values of parameters $q$.

Let us first analyze the behaviour of the slopes $q(x)$ in the vicinity of
the pseudo-crossing in the two-level approximation. Consider two levels
$E_1$ and $E_2$ which cross at $x=x_c$:
\begin{subequations}
\begin{eqnarray}
\label{t2a}
  E_1 &=& q_1(x-x_c), \\
\label{t2b}
  E_2 &=& q_2(x-x_c).
\end{eqnarray}
\end{subequations}
If the interaction matrix element between these two levels is $V$, the
exact adiabatic levels will be
\begin{eqnarray}
\label{t3}
  E_{a,b} &=& \frac{1}{2}\biggl((q_1+q_2)(x-x_c)
\nonumber \\
  &\pm& \sqrt{(q_1-q_2)^2(x-x_c)^2+4V^2}\biggr).
\end{eqnarray}
It is easy now to calculate the energy derivative in respect to $x$
in terms of the mixing angle $\phi$ between unperturbed states 1 and 2:
\begin{eqnarray}
\label{t5}
  \frac{\partial E_{a,b}}{\partial x}
  &=& \cos^2\! \phi\; q_{1,2} + \sin^2\! \phi\; q_{2,1}.
\end{eqnarray}

Note, that at the crossing the angle $\phi$ varies from 0 on one side
through $\pi/4$ in the centre to $\pi/2$ on the other side, which leads to
the change of the slope $q_a(x)=\partial E_a/\partial x$ from $q_1$ through
$(q_1+q_2)/2$ to $q_2$. The narrow crossings with small $V$ are
particularly dangerous, as the slopes change very rapidly within the
interval $\Delta x \approx V/|q_1-q_2|$. Then, even small errors in the
position of the crossing point $x_c$, or the value of $V$ can cause large
errors in $q_{a,b}$. In this model we assume that nondiagonal term $V=
\textrm{const}$. For the real atom $V \propto \alpha^2$. However, if the
crossing region $\Delta x \ll 1$, we can neglect the dependence of $V$ on
$\alpha$.

\paragraph{Semiempirical treatment of the strong interaction of levels:
multi-level case.} \Eref{t5} can be easily generalized to a multi-level
case as it simply gives the slope of a physical level $a$ as a weighted
average of the mixed levels. Thus, if the level $a$ can be expressed as a
linear combination of some unperturbed $LS$-states $\psi_{L_n,S_n}$:
\begin{eqnarray}
\label{t6}
  |a\rangle &=& \sum_n C_n |\psi_{L_n,S_n}\rangle,
\end{eqnarray}
the resultant slope $q_a$ is given by:
\begin{eqnarray}
\label{t7}
  q_a &=& \sum_n C_n^2\; q_n.
\end{eqnarray}
Here again we neglect weak dependence of interaction $V$ on $x$ in
comparison to strong dependence of $C_n^2$ on $x$ near crossing points.

\Eref{t7} allows to improve \textit{ab initio} coefficients $q$ if we can
find the expansion coefficients $C_n$ in \Eref{t6}. That can be done, for
example, by fitting $g$-factors. The magnetic moment operator $\bm{\mu} =
g_0(\bm{L}+2\bm{S})$ is diagonal in $L$ and $S$ and, therefore, does not
mix different $LS$-states. Thus, in the $LS$-basis the resultant $g$-factor
for the state $a$ has exactly the same form as $q_a$:
\begin{eqnarray}
\label{t8}
  g_a &=& \sum_n C_n^2\; g_n.
\end{eqnarray}
If the experimental $g$-factors are known, one can use \Eref{t8} to find
weights $C_n^2$ and, then find the corrected values of the slopes $q_a$.

Sometimes, the experimental data on g-factors are incomplete. Than, one can
still use a simplified version of Eqs.~\eref{t7} and~\eref{t8}:
\begin{subequations}
\label{t9}
\begin{eqnarray}
  g_a &=& C^2 g_a^0 + (1-C^2)\; \bar{g},\;
  \Rightarrow \;
  C^2 = \frac{g_a-\bar{g}}{g_a^0-\bar{g}},\\
  q_a &=& C^2 q_a^0 + (1-C^2)\; \bar{q}.
\end{eqnarray}
\end{subequations}
$C^2$ here is the weight of the dominant $LS$-level in the
experimental one, and the bar means the averaging over the admixing levels.
Of course, there is some arbitrariness in calculation of averages $\bar{g}$
and $\bar{q}$. However, the advantage of Eqs.~\eref{t9} is that only one
experimental $g$-factor is required.

\section{Details of the calculation and results}
\label{details}

As we mentioned above, we performed calculations of energy levels for three
values of the parameter $x$: $x_-=-1/8$, $x_0=0$, and $x_+=1/8$. All three
calculations were done at exactly same level of approximation, to minimize
the error caused by the incompleteness of the basis sets and configuration
sets. From these calculations we found two approximations for $q$:
$q_-=8(\omega(x_0)-\omega(x_-))$ and $q_+=8(\omega(x_+)-\omega(x_0))$. If
there were problems with level identification we performed additional
calculation for $x=0.01$, where the $LS$-coupling should be very accurate
and identification is straitforward. The noticeable difference between
$q_-$ and $q_+$ signaled the possibility of the level crossing. In these
cases we applied the semiempirical procedure described in Sec.~\ref{theory}
to find the corrected values for $q$; otherwise, we simply took the
average: $q=(q_+ +q_-)/2$.

\subsection {Zn II}
\label{Zn}

Zn II has the ground state configuration $[1s^2 \dots 3d^{10}]4s$ and we
are interested in the $4s \rtw 4p_j$ transitions. As the theory here is
much simpler than for other ions, we used Zn II to study the importance of
the core-valence correlation correction and Breit correction to the slopes
$q$. The former correction was calculated in Brueckner approximation:
\begin{eqnarray}
\label{brkr}
  \left(H_{\textrm{DHF}} + \Sigma(E)\right)\Psi &=& E\Psi,
\end{eqnarray}
with the self-energy operator $\Sigma(E)$ calculated in the second order of
MBPT (the perturbation here is the difference between the exact and
Dirac-Hartree-Fock Hamiltonians, $V=H-H_{\textrm{DHF}}$).
The $H_{\textrm{DHF}}$ was calculated with the magnetic part of the Breit
operator included self-consistently. The retardation part of the Breit
operator is known to be significantly smaller \cite{LMY89} and we
completely neglected it here.

The results of our calculations of the frequencies $\omega$ and the slopes
$q$ for two transitions $4s \rtw 4p_j$, $j=1/2,3/2$ are given in
\tref{tab_Zn}. One can see, that both Brueckner-Coulomb and
Brueckner-Coulomb-Breit approximations give very good transition
frequencies, accurate to 0.2\%, though the latter slightly underestimates
the fine splitting. Breit correction to the parameters $q$ does not exceed
1\%, while core-valence correlations account for the 17\% correction.

\begin{table}[tb]
\caption{Transition frequencies and parameters $q$ for Zn~II (in~\cm).
Calculations were done in four different approximations:
Dirac-Hartree-Fock-Coulomb (DHFC), Dirac-Hartree-Fock-Coulomb-Breit
(DHFCB), Brueckner-Coulomb (BC), and Brueckner-Coulomb-Breit (BCB).}

\label{tab_Zn}

\begin{tabular}{lclccccc}
\hline
\hline
\multicolumn{3}{c}{Transition}
&\multicolumn{1}{c}{Exper.}
&\multicolumn{1}{c}{DHFC}
&\multicolumn{1}{c}{DHFCB}
&\multicolumn{1}{c}{BC}
&\multicolumn{1}{c}{BCB}
\\
\hline
&\multicolumn{7}{c}{transition frequencies}\\
 $4s_{1/2}$& $\rtw$&$4p_{1/2}$
 & 48481.077 & 44610.1 & 44608.1 & 48391.2 & 48389.4 \\
           & $\rtw$&$4p_{3/2}$
 & 49355.002 & 45346.9 & 45330.0 & 49263.8 & 49244.6 \\
\hline
&\multicolumn{7}{c}{parameters $q=(q_+ + q_-)/2$}\\

 $4s_{1/2}$& $\rtw$&$4p_{1/2}$&& 1362 & 1359 & 1594 & 1590 \\
           & $\rtw$&$4p_{3/2}$&& 2129 & 2109 & 2500 & 2479 \\
\hline
\hline
\end{tabular}
\end{table}


In \tref{tab_Zn} we did not give separately the values of $q_\pm$. The
difference between them is close to 1\%.
Indeed, in the absence of close interacting levels the dependence of $q$ on
$x$ arise from the corrections to the energy of the order of $\alpha^4
Z^4$, which are very small.

\subsection {Si II}
\label{Si}

Si II has three valence electrons and the ground state configuration $[1s^2
\dots 2p^{6}]3s^2 3p$. Excited configurations of interest are $3s3p^2$ and
$3s^2 4s$. We made the CI calculation in the Coulomb approximation on the
basis set, which included $1s-8s$, $2p-8p$, $3d-8d$, and $4f,\;5f$
orbitals, which we denote as the basis set [8spd5f]. Note, that we use
virtual orbitals, which are localized within the atom \cite{Bv83}, rather
than Dirac-Fock ones. This provides fast convergence. CI included all
single-double (SD) and partly triple excitations from three valence
configurations listed above. The results of these calculations are given in
\tref{tab_Si}.

\begin{table}[tb]
\caption{Transition frequencies $\omega$ from the ground state
$^2\!P_{1/2}^o$, fine structure splitting $\Delta_{\textrm{FS}}$, and
parameters $q_\pm$ for Si~II (in~\cm).}

\label{tab_Si}

\begin{tabular}{lrrrrrr}
\hline
\hline
&\multicolumn{2}{c}{Experiment \cite{Moo58}}
&\multicolumn{4}{c}{Theory}\\
&\multicolumn{1}{c}{$\omega$}
&\multicolumn{1}{c}{$\Delta_{\textrm{FS}}$}
&\multicolumn{1}{c}{$\omega$}
&\multicolumn{1}{c}{$\Delta_{\textrm{FS}}$}
&\multicolumn{1}{c}{$\quad q_-$}
&\multicolumn{1}{c}{$\quad q_+$}\\
\hline
$^2\!P_{3/2}^o$&   287 &\quad 287 &   293 &\quad 293 &\quad 295 &\quad 291 \\
$^4\!P_{1/2}$  & 44080 &     &\quad 41643 &          &\quad 453 &\quad 451 \\
$^4\!P_{3/2}$  & 44191 &\quad 111 & 41754 &      111 &\quad 565 &\quad 564 \\
$^4\!P_{5/2}\;$& 44364 &\quad 174 & 41935 &      181 &\quad 746 &\quad 744 \\
$^2\!D_{3/2}$  & 55304 &\quad     & 54655 &          &\quad 509 &\quad 507 \\
$^2\!D_{5/2}$  & 55320 &\quad  16 & 54675 &       20 &\quad 530 &\quad 530 \\
$^2\!S_{1/2}$  & 65495 &\quad     & 65148 &          &\quad  40 &\quad  39 \\
\hline
\hline
\end{tabular}
\end{table}

Like in Zn, the left and write derivatives $q_-$ and $q_+$ are close to
each other, and all levels with equal exact quantum numbers are well
separated. The astrophysical data exist for the levels $^2\!S_{1/2}$ and
$^2\!D_{5/2}$. The former corresponds to the $3p \rtw 4s$ transition and
has small slope $q$, while the latter corresponds to the $3s \rtw 3p$
transition and has much larger positive $q$. That is in agreement with the
fact, that relativistic corrections to the energy usually decrease with the
principle quantum number $n$ and with the orbital quantum number $l$.
Therefore, for the $ns \rtw np$ transition one should expect large and
positive $q$, while for $np \rtw (n+1)s$, there should be large
cancellation of relativistic corrections to upper and to lower levels,
resulting in smaller $q$ (see disscussion in \cite{DFW99a,DFW99b}). The
dominant correction to our results should be from the core-valence
correlations. In the recent calculations of Mg, which has the same core as
Si~II, the core-valence corrections to transition frequencies were found to
be about 4\% \cite{PKR00b,PKRD01}. We conservatively estimate corresponding
correction to $q$ to be 6\% of the larger $q$, i.e. 30~\cm.

\subsection {Cr II}
\label{Cr}

Cr II has the ground state configuration $[1s^2
\dots 3p^{6}]3d^5$ with five valence electrons. The astrophysical data
correspond to the $3d \rtw 4p$ transition, for which one may expect
negative value of $q$. CI calculations here are much more complicated, than
for Si~II. There is strong relaxation of the $3d$ shell in the discussed
transition, which requires more basic $d$-orbitals. Therefore, we used the
[6sp9d6f] basis set.
In CI we included only single and double (SD) excitations. Some of the
triple, quadruple, and octuple excitations were accounted for by means of
the second order perturbation theory. It was found that corresponding
corrections to transition frequencies were of the order of few percent, and
were even smaller for parameters $q$. In general, these corrections did not
improve the agreement with the experiment, so we present only CI results in
\tref{tab_Cr}.

\begin{table}[tb]
\caption{Transition frequencies $\omega$ from the ground state
$^6\!S_{5/2}$, fine structure splitting $\Delta_{\textrm{FS}}$, and
parameters $q$
 for Cr~II (in~\cm). CI single-double approximation
was used for the Coulomb-Breit interaction.}

\label{tab_Cr}

\begin{tabular}{lrrrrr}
\hline
\hline
&\multicolumn{2}{c}{Experiment}
&\multicolumn{3}{c}{Theory}\\
&\multicolumn{1}{c}{$\omega$}
&\multicolumn{1}{c}{$\Delta_{\textrm{FS}}$}
&\multicolumn{1}{c}{$\omega$}
&\multicolumn{1}{c}{$\Delta_{\textrm{FS}}$}
&\multicolumn{1}{c}{$\quad q_+$}
\\
\hline
$^6\!D_{5/2}$    & 12148 &\quad     & 13123 &\quad     &\quad $-2314$ \\
$^6\!D_{7/2}$    & 12304 &\quad 156 & 13289 &\quad 165 &\quad $-2153$ \\

$^6\!F_{1/2}^o$  & 46824 &\quad     & 47163 &\quad     &\quad $-1798$ \\
$^6\!F_{3/2}^o$  & 46906 &\quad  82 & 47244 &\quad  81 &\quad $-1715$ \\
$^6\!F_{5/2}^o$  & 47041 &\quad 135 & 47378 &\quad 134 &\quad $-1579$ \\
$^6\!F_{7/2}^o$  & 47228 &\quad 187 & 47565 &\quad 187 &\quad $-1387$ \\
$^6\!F_{9/2}^o$  & 47465 &\quad 237 & 47803 &\quad 238 &\quad $-1148$ \\
$^6\!F_{11/2}^o$ & 47752 &\quad 287 & 48091 &\quad 288 &\quad $ -862$ \\

$^6\!P_{1/2}^o $ & 48399 &\quad     & 48684 &\quad     &\quad $-1364$ \\
$^6\!P_{2/2}^o $ & 48491 &\quad  92 & 48790 &\quad 106 &\quad $-1278$ \\
$^6\!P_{3/2}^o $ & 48632 &\quad 141 & 48947 &\quad 157 &\quad $-1108$ \\
\hline
\hline
\end{tabular}
\end{table}

As we mentioned above, there is strong relaxation of the $3d$-shell in the
$3d \rtw 4p$ transition. We were not able to saturate CI space and
completely account for this relaxation. Because of that, we estimate the
error for $q$ here to be close to 10\%.

We have seen before for Zn II and Si II, that in the absence of
level-crossing the difference between $q_+$ and $q_-$ is smaller than other
theoretical uncertainties. In Cr II there are no close levels which may
interact with each other, so in the calculation presented in \tref{tab_Cr}
we determined only the right derivative $q_+$. In calculations with
different basis sets we checked that the difference between $q_+$ and $q_-$
is much smaller than the given above theoretical error (see
\tref{tab_fin}).

\subsection {Fe II}
\label{Fe}

Fe II ion has 7 valence electrons in configuration $3d^64s$ and represents
the most complicated case. The astrophysical data includes 5 lines in the
band 38000~\cm --~43000~\cm and two lines with the frequency close to
62000~\cm. The first band consists of three close, but separated multiplets
with a regular fine structure splittings. The 62000~\cm band is completely
different as the multiplets here strongly overlap and fine structure
intervals are irregular \cite{Moo58}. Characteristic distance between the
levels with identical exact quantum numbers is few hundred~\cm, which is
comparable to the fine structure splittings. This means that the levels
strongly interact and even their identification may be a problem.

In fact, in Moore Tables \cite{Moo58} one of the multiplets of interest,
namely $y\,^6\!P^o$, is erroneously assign to the configuration
$3d^6(^7\!S)4p$. It is an obvious misprint, as there is no term $^7\!S$ for
configuration $3d^6$. This term appears, however, in the configuration
$3d^5$ and the correct assignment of this multiplet should be
$3d^5(^7\!S)4s4p$. This assignment is in agreement with our calculations
and with the experimental $g$-factor of the level with $J=7/2$. We
checked that all close levels of the configuration $3d^64p$ have
significantly smaller $g$-factors.

\begin{figure}[tb]
\includegraphics[scale=0.4]{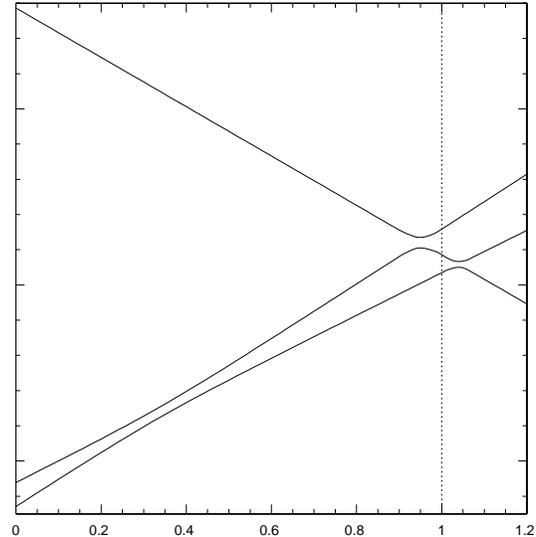}
\caption{Examples of typical interaction of levels in the upper band of Fe~II.
Levels are shown in arbitrary units as function of
$(\alpha/\alpha_0)^2=x+1$. Levels of configuration $3d^6 4p$ have similar
slopes and strongly interact with each other. That causes wide
pseudo-crossings, similar to one shown on the left side of the plot. The
level $^6\!P_{7/2}^o$ of the configuration $3d^5 4s4p$ moves in the
opposite direction. A series of sharp pseudo-crossings takes place near the
physical value of $\alpha$, marked by a vertical dotted line.}
\label{fig_fe}
\end{figure}

This reassignment has dramatic consequences in terms of the corresponding
parameter $q$ as configurations $3d^64p$ ($4s-4p$ transition from the
ground state) and $3d^54s4p$ ($3d-4p$ transition) move in the opposite
directions from the ground state configuration $3d^64s$ when $x$ is
changed. It also causes a number of pseudo-crossings to occur right in the
vicinity of $x=0$ (see Fig.~\ref{fig_fe}).

CI calculations for Fe II were done on the basis set [6spdf] in the SD
approximation (see \tref{tab_Fe}). Triple excitations were included within
second order perturbation theory and corresponding corrections were found
to be relatively small. One can see from \tref{tab_Fe} that for the lower
band both frequencies and $g$-factors are reproduced rather accurately.

\begin{table}[tb]
\caption{Transition frequencies $\omega$ from the ground state
$^6\!D_{9/2}$, $g$-factors, and
parameters $q_\pm$ for Fe~II (in~\cm).}

\label{tab_Fe}

\begin{tabular}{lllrrrrr}
\hline
\hline
&\multicolumn{2}{c}{Experiment}
&\multicolumn{4}{c}{Theory}\\
&\multicolumn{1}{c}{$\omega$}
&\multicolumn{1}{c}{$g$}
&\multicolumn{1}{c}{$\omega$}
&\multicolumn{1}{c}{$g$}
&\multicolumn{1}{c}{$g(LS)$}
&\multicolumn{1}{c}{$\quad q_-$}
&\multicolumn{1}{c}{$\quad q_+$}
\\
\hline
$^6\!D_{9/2}^o$ &38459\;& 1.542 &\;38352&       & 1.556 & $ 1359 $&$ 1363$\\
$^6\!D_{7/2}^o$ & 38660 & 1.584 & 38554 & 1.586 & 1.587 & $ 1522 $&$ 1510$\\

$^6\!F_{11/2}^o$& 41968 &       & 41864 &       & 1.455 & $ 1496 $&$ 1508$\\
$^6\!F_{9/2}^o$ & 42115 & 1.43  & 42012 &       & 1.434 & $ 1615 $&$ 1631$\\
$^6\!F_{7/2}^o$ & 42237 & 1.399 & 42141 & 1.396 & 1.397 & $ 1738 $&$ 1737$\\

$^6\!P_{7/2}^o$ & 42658 & 1.702 & 42715 & 1.709 & 1.714 &$ 1241 $&$ 1261$ \\
$^4\!D_{7/2}^o$ & 44447 & 1.40  & 44600 & 1.345 & 1.429 &$ 1791 $&$ 1837$ \\
$^4\!F_{7/2}^o$ & 44754 & 1.29  & 44386 & 1.327 & 1.238 &$ 1608 $&$ 1601$ \\

$^8\!P_{7/2}^o$ & 54490 &       & 54914 & 1.936 & 1.937 &$-2084 $&$-2086$ \\

$^4\!G_{7/2}^o$ & 60957 & 0.969 & 63624 & 0.978 & 0.984 &$ 1640 $&$ 1640$ \\
$^4\!H_{7/2}^o$ & 61157 & 0.720 & 63498 & 0.703 & 0.667 &$ 1296 $&$ 1247$ \\
$^4\!D_{7/2}^o$ & 61726 & 1.411 & 66145 & 1.398 & 1.429 &$ 1194 $&$ 1240$ \\
$^4\!F_{7/2}^o$ & 62066 & 1.198 & 65528 & 1.252 & 1.238 &$ 1071 $&$ 1052$ \\
$^6\!P_{7/2}^o$ & 62172 & 1.68  & 65750 & 1.713 & 1.714 &$-1524 $&$-1514$ \\
$^2\!G_{7/2}^o$ & 62323 &       & 64798 & 0.882 & 0.889 &$ 1622 $&$ 1605$ \\
\hline
\hline
\end{tabular}
\end{table}

The first anomaly takes place at 44000~\cm, where the levels
$^4\!D_{7/2}^o$ and $^4\!F_{7/2}^o$ appear in the reverse order.
Theoretical $g$-factors are also much further from $LS$~values (1.429 and
1.238). That means that theoretical levels are at pseudo-crossing, while
experimental levels already passed it. Indeed, calculations for $x=1/8$
show that the right order of levels is restored, though the $g$-factors are
still too far from $LS$~values.

The second anomaly corresponds to the band above 60000~\cm. Here the order
of calculated levels differs from that of the experimental ones. Note, that
for this band only levels of negative parity with $J=7/2$ are given in
\tref{tab_Fe}. Thus, all of them can interact with each other. Let us
estimate, how this interaction can affect the slopes $q$.

Five levels from this band belong to configuration $3d^6 4p$ and have close
slopes with the average $\bar{q}=1360$~\cm. Only the level $^4\!F^o_{7/2}$
has the slope, which is 300~\cm\ smaller, than the average. The remaining
level $^6\!P^o_{7/2}$ belongs to configuration $3d^5 4s4p$ and has the
slope of the opposite sign $q_1=-1519$~\cm. Its absolute value is 500~\cm\
smaller, than for the level $^8\!P^o_{7/2}$ of the same configuration
$3d^5 4s4p$. That
suggests that the levels $^4\!F^o_{7/2}$ and $^6\!P^o_{7/2}$ strongly
interact with each other. This is also in agreement with the fact, that
these levels are the closest neighbors both experimentally and
theoretically and that they cross somewhere between $x_-$ and $x$.
There is also strong interaction between the levels $^2\!G^o_{7/2}$,
$^4\!F^o_{7/2}$, and $^4\!D^o_{7/2}$. That can be seen if one calculates
the scalar products (overlaps)
 between corresponding wave functions for different
values of $x$, such as: $\langle i(x_-)|k(x_+)\rangle$. For weekly
interacting levels $\langle i(x_-)|k(x_0)\rangle \approx \langle
i(x_-)|k(x_+)\rangle \approx \delta_{i,k}$, so large non-diagonal matrix
elements signal, that corresponding levels interact.

Interaction of levels $^2\!G^o_{7/2}$, $^4\!F^o_{7/2}$, and $^4\!D^o_{7/2}$
does not affect the slopes $q$ as strongly, as the interaction of $^4\!F^o_{7/2}$
and $^6\!P^o_{7/2}$, so we can account for the former in a less accurate
way, but it is important to include the latter as accurately as possible.

The level $^6\!P^o_{7/2}$ interacts with some linear
combination of levels $^2\!G^o_{7/2}$, $^4\!F^o_{7/2}$, and
$^4\!D^o_{7/2}$. The slopes and $g$-factors of the latter are relatively
close to each other, so we can simply take the average for all three:
\begin{eqnarray}
\label{f1}
  \bar{g} = 1.185; \quad \bar{q} = 1297.
\end{eqnarray}
Now we can use experimental $g$-factor of the state $^6\!P^o_{7/2}$ and
\Eref{t9} to determine the mixing:
\begin{eqnarray}
\label{f2}
  &&C^2 = \frac{1.68-\bar{g}}{1.713-\bar{g}}=0.937,\\
\label{f3}
  &&q(^6\!P^o_{7/2}) =
-1342.
\end{eqnarray}
\Eref{f3} corresponds to the correction $\delta q=+177$. Therefore, for
the closest level $^4\!F^o_{7/2}$ this model gives an estimate:
\begin{eqnarray}
\label{f4}
  &&q(^4\!F^o_{7/2}) = \bar{q} - \delta q  = 1120.
\end{eqnarray}

Eqs.~\eref{f3} and \eref{f4} show that correction for the mixing is not
very large. That corresponds to the fact that experimental $g$-factor of the
level $^6\!P^o_{7/2}$ is significantly larger than any $g$-factors of the
levels of the configuration $3d^6 4p$. Thus, the interaction for this level
is relatively small. On the contrary, the levels of the configuration $3d^6
4p$ strongly interact with each other, but corresponding changes of the
slopes are also relatively small (since the $q$ values for these
strongly interacting levels are approximately the same).

We estimate the accuracy of our calculations for the lower band of Fe~II to
be about 150~\cm, and approximately 300~\cm\ for the values~\eref{f3} and
\eref{f4}.

\subsection {Ni II}
\label{Ni}

\begin{figure}[tb]
\includegraphics[scale=0.4]{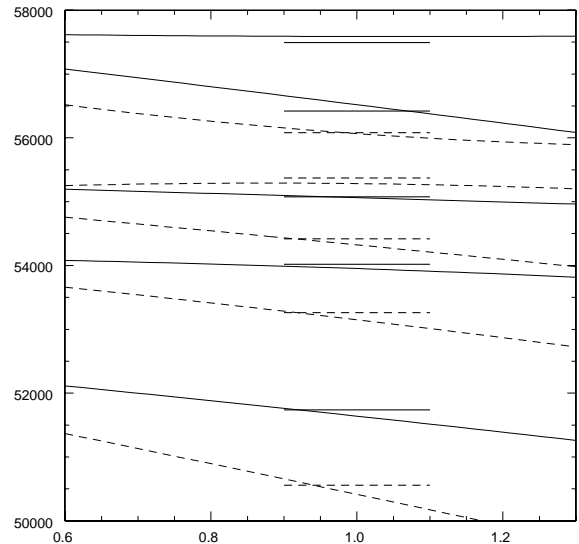}
\caption{Dependence of the odd levels of Ni II on
$(\alpha/\alpha_0)^2=x+1$. Solid lines correspond to $J=5/2$ and dashed
lines to $J=7/2$. The experimental positions of the lines are shown as
short horizontal lines and are all shifted by 1000~\cm. The order of levels
from bottom up: $^4\!D_{7/2,5/2}^o$, $^4\!G_{7/2,5/2}^o$,
$^4\!F_{7/2,5/2}^o$, $^2\!G_{7/2}^o$, $^2\!F_{5/2}^o$, $^2\!D_{5/2}^o$,
and $^2\!F_{5/2}^o$.}
\label{fig1}
\end{figure}

Ni II has the ground state configuration $3d^9$. The spectrum is somewhat
simpler, than for Fe~II. There are also pseudo-crossings here, but they
either lie far from $x=0$, or are rather wide. That makes their treatment
slightly easier. Nevertheless, our results significantly differ from
previous calculations \cite{DFM01}.

CI calculations were done for the Coulomb potential and included SD and
partly triple excitations on the basis set [5spdf]. We calculated 5 lower
odd levels with $J=5/2$ and 5 with $J=7/2$ for $x_-,\,x_0$, and $x_+$, and
used parabolic extrapolation for the interval $-0.4\le x \le +0.3$ (see
Fig.~\ref{fig1}). It is seen that the theory accurately reproduce relative
positions of all levels. An overall agreement between the theory and the
experiment becomes close to perfect if all experimental levels are shifted
by 1000~\cm\ down, as it is done in Fig.~\ref{fig1}. Note, that this shift
constitutes only 2\% of the average transition frequency.

\begin{table}[tb]
\caption{Transition frequencies $\omega$ from the ground state
$^2\!D_{5/2}$, $g$-factors, and
parameters $q_\pm$ for Ni~II (in~\cm).}

\label{tab_Ni}

\begin{tabular}{lrlrrrrr}
\hline
\hline
&\multicolumn{2}{c}{Experiment}
&\multicolumn{4}{c}{Theory}\\
&\multicolumn{1}{c}{$\omega$}
&\multicolumn{1}{c}{$g$}
&\multicolumn{1}{c}{$\omega$}
&\multicolumn{1}{c}{$g$}
&\multicolumn{1}{c}{$g(LS)$}
&\multicolumn{1}{c}{$\quad q_-$}
&\multicolumn{1}{c}{$\quad q_+$}
\\
\hline
$^2\!D_{3/2}  $ &  1507 &       &  1579 &       & 0.800 & $ 1559 $&$ 1552$\\

$^4\!D_{7/2}^o$ & 51558 &\;1.420 & 50415 & 1.423 & 1.429 & $-2405 $&$-2425$\\
$^4\!D_{5/2}^o$ &\;52739&\;1.356 &\;51640&\;1.360& 1.371 & $-1217 $&$-1245$\\

$^4\!G_{7/2}^o$ & 54263 &\;1.02  & 53150 & 1.016 & 0.984 & $-1334 $&$-1387$\\
$^4\!G_{5/2}^o$ & 55019 &\;0.616 & 53953 & 0.617 & 0.571 & $ -370 $&$ -418$\\

$^4\!F_{7/2}^o$ & 55418 &\;1.184 & 54323 & 1.183 & 1.238 & $-1104 $&$-1124$\\
$^4\!F_{5/2}^o$ & 56075 &\;0.985 & 55063 & 0.986 & 1.029 & $ -332 $&$ -334$\\

$^2\!G_{7/2}^o$ & 56372 &\;0.940 & 55284 & 0.933 & 0.889 & $  -60 $&$ -188$\\

$^2\!F_{7/2}^o$ & 57080 &\;1.154 & 56067 & 1.128 & 1.143 & $ -911 $&$ -713$\\
$^2\!D_{5/2}^o$ & 57420 &\;1.116 & 56520 & 1.108 & 1.200 & $-1419 $&$-1438$\\
$^2\!F_{5/2}^o$ & 58493 &\;0.946 & 57589 & 0.959 & 0.857 & $   -35$&$   -5$\\
\hline
\hline
\end{tabular}
\end{table}

Calculated $g$-factors are generally in agreement with the
experiment~\cite{Moo58} and noticeably different from the pure $LS$-values
(see \tref{tab_Ni}). However, for the level $^2\!F_{7/2}^o$ theoretical
$g$-factor is smaller than the $LS$ value, while experimental one is
larger. There are no nearby levels who may mix to this one and move
$g$-factor closer to experiment. On the other hand, the difference with
experiment is only 2\% and may be within experimental accuracy.

Fig.~\ref{fig1} shows that the levels $^2\!G_{7/2}^o$ and $^2\!F_{7/2}^o$
cross at $x \approx 0.3$ and they already strongly interact at $x=0$.
Theoretical splitting for these levels is 10\% larger than experimental
one. Thus, they are in fact even closer to the crossing point than is
predicted by the theory. The experimental splitting is equal to the
theoretical one for larger value of $\alpha$ corresponding to $x
\approx 0.15$. At $x=0.15$ the slopes of these levels are $-265$ and $-590$,
and for $x=0$ they are $-124$ and $-812$ correspondingly. Note, that the
sum of the slopes at $x=0.15$ differs by 80~\cm\ from the sum at $x=0$.
According to \Eref{t5} for a two-level system the sum is constant. This
means that these two levels repel from the lower lying level
$^4\!F_{7/2}^o$. Taking this analysis into account we suggest an average
between $x=0$ and $x=0.15$ as our final value:
$q(^2\!F_{7/2}^o)=-700(250)$.

\subsection*{Conclusions}
\label{discussion}

In this paper we present new refined calculations of the parameters $q$,
which determine $\alpha$-dependence of the transition frequencies for a
number of ions used in the astrophysical search for $\alpha$-variation.
These ions appear to be very different from the theoretical point of view.
Because of that we had to use different methods and different levels of
approximation for them. The final accuracy of our results differs not only
for different ions, but also for different transitions.

The simplest system is Zn~II, which has one valence electron. On the other
hand, this is the heaviest ion and it has the largest core, which includes
$3d^{10}$-shell. That gave us the opportunity to study corrections to $q$
from the core-valence correlations and from Breit interaction. We found the
former to be about 17\% and the latter to be less than 1\%. For lighter
ions Breit interaction should be even smaller and can be safely neglected.
Other ions also have much smaller and more rigid cores, so one might expect
that core-valence correlations are few times weaker there in comparison to
Zn. That allows us to neglect core-valence correlations for all other ions
discussed in this paper.

Si~II has the smallest core $1s^2\dots 2p^6$ and three valence electrons.
For neutral Mg, which has the same core, the core-valence corrections to
the $3s \rtw 3p$ transition frequencies were found to be about 4\%
\cite{PKR00b,PKRD01}. CI calculation for Si~II is relatively simple and the
errors associated with incompleteness of CI space are small. Thus, our
estimate of the accuracy for Si on 6\% level seems to be rather
conservative.

Cr, Fe, and Ni have the core $1s^2\dots 3p^6$ and the core excitation
energy varies from 2~a.u. for Cr~II to 2.6~a.u. for Ni~II. In comparison,
the core excitation energy for Zn~II is 0.9~a.u. Therefore, we estimate the
core-valence correlation corrections for these ions to be at least two
times smaller, than for Zn~II.

Additional error here is associated with incompleteness of the CI space.
These ions have from 5 to 9 valence electrons and CI space can not be
saturated. To estimate corresponding uncertainty we performed several
calculations for each ion using different basis sets and two different
computer packages described in Sec.~\ref{theory}. The basic
Dirac-Hartree-Fock orbitals were calculated for different configurations
(for example, for the ground state configuration and for excited state
configuration, etc.).

Supplementary information on the accuracy of our calculations can be
obtained from comparison of calculated spectra and $g$-factors with
experimental values. The later appear to be very important as they give
information about electron coupling, which depends on relativistic
corrections and on interaction between $LS$-multiplets. Our results for
Cr~II appear to be very close for different calculations and are in good
agreement with the experiment both in terms of the gross level structure
and spin-orbit splittings (see \tref{tab_Cr}), so we estimate our final
error here to be about 10~--~12\%.

The largest theoretical uncertainties appear for Fe~II and Ni~II where the
number of valence electrons is largest and the interaction of levels is
strongest. Here we had to include semi-empirical fits to improve the
agreement between the theory and the experiment. We took into account the
size of these semi-empirical corrections in estimates of the accuracy of
the calculated values of $q$.

The final results are presented in \tref{tab_fin}. Note again, that they
are based on several independent calculations performed using two different
computer codes. Some of the intermediate results are given in \tref{tab_Zn}
--~\tref{tab_Ni}.

\begin{acknowledgments}
This work is supported by Australian Research Council. One of us (MK)
thanks UNSW for hospitality and acknowledges support from the Gordon
Godfrey Fund.
\end{acknowledgments}


\end{document}